\documentclass[aps,pra,showpacs,twocolumn,amsmath,amssymb,superscriptaddress,footinbib]{revtex4}

\usepackage[english]{babel}
\usepackage{latexsym}
\usepackage{graphicx}
\usepackage{subfigure}
\usepackage{epsfig}
\usepackage{amsfonts}
\usepackage{amssymb}
\usepackage{amsmath}
\usepackage{bbm}

\begin{document}

\title{Multiple time-scale Landau-Zener transitions in many-body systems}


\author{Jonas Larson}
\email{jolarson@physto.su.se} \affiliation{Department of Physics,
Stockholm University, AlbaNova University Center, Se-106 91 Stockholm,
Sweden}
\affiliation{Institut f\"{u}r Theoretische Physik, Universit\"{a}t zu
  K\"{o}ln, De-50937 K\"{o}ln, Germany}

\date{\today}

\begin{abstract}
Motivated by recent cold atom experiments in optical lattices, we consider a lattice version of the Landau-Zener problem. Every single site is described by a Landau-Zener problem, but due to particle tunnelling between neighboring lattice sites this onsite single particle Landau-Zener dynamics couples to the particle motion within the lattice. The lattice, apart from having a dephasing effect on single site Landau-Zener transitions, also implies, in the presence of a confining trap, an inter-site particle flow induced by the Landau-Zener sweeping. This gives rise to an interplay between intra- and inter-site dynamics. The adiabaticity constrain is therefor not simply given by the standard one; the Hamiltonian rate of change relative to the gap of the onsite problem. In experimentally realistic situations, the full system evolution is well described by Franck-Condon physics, e.g. non-adiabatic excitations are predominantly external ones characterized by large phononic vibrations in the atomic cloud, while internal excitations are very weak as close to perfect onsite transitions take place.    
\end{abstract}

\maketitle

\section{Introduction}\label{sec1}
The adiabatic theorem tells us that non-adiabatic excitations become important whenever the rate of change of the Hamiltonian is large compare to the splitting between nearby energies. A typical situation when this happens is when two weakly coupled diabatic, or bare, energy levels cross. The coupling between the corresponding states implies a lifting of the degeneracy and the crossing becomes avoided. The simplest description of this scenario is captured by the Landau-Zener (LZ) model. Being analytically solvable, the LZ formula gives an expression for the transition probability when the system is swept through an avoided crossing~\cite{lz}. In the diabatic representation, the state vector $|\psi(t)\rangle=[\psi_x(t)\,\, \psi_y(t)]^T$ is a solution of the time-dependent Schr\"odinger equation ($\hbar=1$)
\begin{equation}
i\frac{\partial}{\partial t}\left[
\begin{array}{c}
\psi_x(t)\\
\psi_y(t)\end{array}\right]=\left[
\begin{array}{cc}
\lambda t & U \\
U & -\lambda t\end{array}\right]\left[
\begin{array}{c}
\psi_x(t)\\
\psi_y(t)\end{array}\right].
\end{equation}
Here, $\lambda$ is the sweep velocity and $U$ the coupling strength between the two diabatic states. For an initial state $\psi_x(-\infty)=1$, $\psi_y(-\infty)=0$, the probability for transfer from the $x$-diabatic state to the $y$-diabatic state at $t=+\infty$ is given by the expression $P=\exp\left(-\Lambda\right)$ with the adiabaticity parameter $\Lambda=\frac{2\pi U^2}{\lambda}$. The adiabatic states are the instantaneous eigenstates of the corresponding Hamiltonian, and for the LZ problem we note that far from the crossing at $t=0$, the diabatic and adiabatic states coincide up to a swapping of the indices. As mentioned, a slow time-change in comparison to the energy gap, i.e. $\lambda\ll|U|$ (in dimensionless units), implies an adiabatic evolution, here seen in the large $\Lambda$-parameter.
 
Since adiabatic breakdown (i.e. population transfer between adiabatic states) occurs predominantly in the close vicinity of the crossing where the diabatic energies are approximately linear, the LZ model has seen numerous applications in all possible fields of physics; in molecular/chemical physics it demonstrates breakdown of the Born-Oppenheimer approximation~\cite{molLZ}, in cold atom physics it can be used to explain the decay of Bloch oscillations~\cite{coldLZ1} or describe the formation of molecules when the gas is driven through a Feshbach resonance~\cite{coldLZ2}, for solid state Josephson junctions LZ physics can be used to analyze transport properties~\cite{jjLZ1} or interference effects~\cite{jjLZ2}, it may also be used to understand critical slowing down and the Kibble-Zurek mechanism appearing when a system is driven through a critical point~\cite{kzLZ}, and it can be employed for state preparation and coherent control~\cite{ccLZ}. This said, with ever refined experimental techniques in especially the AMO community, more complicated situations can be {\it in situ} studied and as a result extended LZ models become relevant. 

In this paper we consider a lattice version of a many-body LZ (MBLZ) problem, namely ultra cold bosonic atoms loaded into the first excited states, $p$-bands, of a square optical lattice~\cite{girvin}. The analysis is carried out at a mean-field level which is not only computationally tractable, but it is also expected to give an accurate picture of current experiments using cold atomic gases in optical lattices. Every single lattice site realizes a non-linear (single particle) LZ system, but tunnelling of atoms from site-to-site makes the full system very complex showing an interplay between intra- and inter-site dynamics. Such coupled dynamics results in a complicated evolution during the LZ sweep. More precisely, there is an intra-site time-scale (indirectly related to the adiabaticity parameter $\Lambda$) which determines the probability for transitions between the two onsite diabatic states, and there is an inter-site time-scale related to the mobility of particles within the lattice. For physically relevant parameters, intra-site adiabaticity is easier to fulfill than inter-site adiabaticity. This can be understood since the intra-site dynamics describes a macroscopic flow of atoms within the lattice. The effect becomes especially clear as the system size is increased, and in fact by taking the thermodynamic limit properly the system hosts an Ising-type quantum phase transition which prevents full adiabatic driving due to the critical slowing down mechanism. The system is particularly interesting since this type of MBLZ problem is reminiscent of Franck-Condon physics which typically can be found in pump-probe experiments in molecular and chemical physics~\cite{fc}. Here, however, is the external degrees of freedom descritized to the lattice sites and apparently the corresponding `pump' (LZ) process is here taking place for a finite time allowing for external evolution to occur during the `pumping'.  
 
The paper is structured in the following way. The system Hamiltonian, and its physical realization in terms of cold $p$-band atoms, is introduced in the following section, where also the idea of the mean-field approach is outlined. Some general remarks on the lattice problem is also given in Sec.~\ref{sec2}. The results for the many-site problem are reported in Sec~\ref{mssec}. To better understand the interplay between intra- and inter-site dynamics we start by analyzing the ground state properties and then turn the attention to the time-dependent MBLZ problem. We conclude with a summary in Sec.~\ref{seccon}.

\section{Model system}\label{sec2}
In this section we introduce the model that supports different internal and external characteristic time-scales. Effective models describing the MBLZ problems discussed in this paper could arise in various physical systems like trapped spinor condensates in optical lattices~\cite{spinor}, ion traps~\cite{trapion}, impurity-BEC systems in double-wells~\cite{jesse}, or atomic condensates occupying higher energy bands of optical lattices~\cite{pband1}. The following derivation considers the last option, more precisely an ultracold gas of bosonic atoms loaded into the $p$-bands (first excited) of a square optical lattice. The reason for this choice is because the two time-scales appear naturally in this system.

The specific system is presented next by deriving the corresponding many-body Hamiltonian. From the many-body Hamiltonian it is straightforward to arrive at the effective model used in this paper by introducing the coherent state ansatz wave-function and from there on obtain the mean-field equations of motion. 

\subsection{Physical system and the quantum many-body Hamiltonian}
We consider bosonic atoms of mass $m$ free to move in a 2D periodic potential. The single-particle Hamiltonian reads
\begin{equation}\label{spham}
\hat{H}_\mathrm{sp}=-\frac{\hbar^2\nabla^2}{2m}+V(\bf{r}'),
\end{equation}
where the potential $V(\mathbf{r}')=V_\mathrm{lat}(\mathbf{r}')+V_\mathrm{trap}(\mathbf{r}')$ with
\begin{equation}
\begin{array}{l}
V_\mathrm{lat}(\mathbf{r}')=\tilde{V}_x\sin(kx')+\tilde{V}_y\sin(ky'),\\ \\
\displaystyle{V_\mathrm{trap}(\mathbf{r}')=\frac{m\tilde{\omega}^2}{2}\left(x'^2+y'^2\right),}
\end{array}
\end{equation}
consists of the optical lattice and the external trapping potentials respectively, with $k$ being the lattice wave number and $\tilde{\omega}$ the trap frequency. The spectrum consists of bands of allowed energies separated by band-gaps. Thus, the eigenenergies and eigenstates can be labeled by a desecrate band index and two continuous quasi momenta. On the isotropic square lattice $V_\mathrm{lat}(\mathbf{r}')$ there is a single lowest ($s$) band and two degenerate first excited ($p$) bands. Throughout this article we will work with dimensionless variables where the recoil energy $E_R=\hbar^2k^2/2m$ sets the energy scale giving a characteristic length $l=k^{-1}$ and time $\tau=\hbar/E_R$. The scaled dimensionless variables/parameters become $x=kx'$, $y=ky'$, $V_x=\tilde{V}_x/E_R$, $V_y=\tilde{V}_y/E_R$, $\omega=\sqrt{2}m\tilde{\omega}/\hbar k^2$. 

For the isotropic 2D lattice, every lattice site hosts two degenerate atomic orbital states, the $p_x$- and $p_y$-orbitals. These orbitals represent the single site atomic states and are given by the Wannier functions. Since we have two of them for every site, we can think of the single site state as a spin-1/2 or qubit particle, but it should be remembered that the spatial dependence of the orbitals is important for giving a full description of the atomic states. Just like the eigenstates of an isotropic 2D harmonic oscillator, the $p_x$-orbital state is roughly Gaussian in the $y$-direction and has a single node in the $x$-direction. Its width is larger in the direction of the node, i.e. the $x$-direction. The same properties apply for the $p_y$-orbital but with the two directions swapped. Naturally, the shapes of these orbitals (Wannier functions) will play an important role for the dynamics of the system. 

From the single particle Hamiltonian (\ref{spham}) we continue by applying the second quantization procedure. The general form of the many-body Hamiltonian is
\begin{equation}\label{mbham}
\hat{H}_\mathrm{mb}=\int\,d\mathbf{r}'\,\hat{\Psi}^\dagger(\mathbf{r}')\left[\hat{H}_\mathrm{sp}+\frac{U}{2}\hat{\Psi}^\dagger(\mathbf{r}')\hat{\Psi}(\mathbf{r}')\right]\hat{\Psi}(\mathbf{r}').
\end{equation}
Here, $U$ is the effective atom-atom ($s$-wave) interaction strength and $\hat{\Psi}(\mathbf{r}')$ and $\hat{\Psi}^\dagger(\mathbf{r}')$ are the atomic annihilation and creation operators respectively which obey the regular bosonic commutation statistics $\left[\hat{\Psi}(\mathbf{r}''),\hat{\Psi}^\dagger(\mathbf{r}')\right]=\delta(\mathbf{r}''-\mathbf{r}')$. The Hamiltonian (\ref{mbham}) is exact within the framework of two-body contact interactions. In order to derive an effective second quantized model we impose the single-band and tight-binding approximations, i.e. we will restrict the atoms to reside only on the two $p$-bands and only consider tunnelling between nearest neighbors as well as only onsite interactions. To this end we expand the atom operators in the $p$-band Wannier functions
\begin{equation}\label{atop}
\hat{\Psi}(\mathbf{r})=\sum_{\alpha\mathbf{j}}w_{\alpha\mathbf{j}}(\mathbf{r})\hat{a}_{\alpha\mathbf{j}},
\end{equation}
where $w_{\alpha\mathbf{j}}(\mathbf{r})$ is the $p_\alpha$-orbital ($\alpha=x,\,y$) Wannier function at site $\mathbf{R}_\mathbf{j}=(x_\mathbf{j},y_\mathbf{j})=(\pi j_x,\pi j_y)$, and $\hat{a}_{\alpha\mathbf{j}}$ annihilates an atom at site $\mathbf{j}$ in orbital $\alpha$. The creation/annihilation operators obey the boson commutation relation $\left[\hat{a}_{\alpha\mathbf{i}},\hat{a}_{\beta\mathbf{j}}^\dagger\right]=\delta_{\alpha\beta}\delta_{\mathbf{ij}}$. Inserting (\ref{atop}), and its hermitian conjugate, in the expression (\ref{mbham}) for the many-body Hamiltonian and make use of the orthogonality of the Wannier functions together with imposing the tight-binding approximation, one derives the second quantized Hamiltonian,
\begin{equation}\label{secondham}  
\hat{H}=\hat{H}_0+\hat{H}_\mathrm{dd}+\hat{H}_\mathrm{oc},
\end{equation}
with 
\begin{equation}\label{kinterm}
\begin{array}{lll}
\hat{H}_0 & = & \displaystyle{-\sum_{\alpha,\beta}\sum_{\langle\mathbf{ij}\rangle_\alpha}t_{\alpha\beta}\hat{a}_{\beta\mathbf{i}}^\dagger\hat{a}_{\beta\mathbf{j}}}\\ \\
& & \displaystyle{+\sum_\alpha\sum_\mathbf{j}\left[E_{\alpha}(t)+\frac{\omega^2}{2}\left(x_\mathbf{j}^2+y_\mathbf{j}^2\right)\right]\hat{n}_{\alpha\mathbf{j}}},
\end{array}
\end{equation}
and the interaction parts
\begin{equation}
\hat{H}_\mathrm{dd}=\frac{U}{2}\sum_\alpha\sum_\mathbf{j}\hat{n}_{\alpha\mathbf{j}}\left(\hat{n}_{\alpha\mathbf{j}}-1\right)+\frac{U}{3}\sum_{\alpha\beta,\alpha\neq\beta}\sum_\mathbf{j}\hat{n}_{\alpha\mathbf{j}}\hat{n}_{\beta\mathbf{j}};
\end{equation}
and 
\begin{equation}\label{ocint}
\hat{H}_\mathrm{oc}=\frac{U}{6}\sum_{\alpha\beta,\alpha\neq\beta}\sum_\mathbf{j}\left(\hat{a}_{\alpha\mathbf{j}}^\dagger\hat{a}_{\alpha\mathbf{j}}^\dagger\hat{a}_{\beta\mathbf{j}}\hat{a}_{\beta\mathbf{j}}+\hat{a}_{\beta\mathbf{j}}^\dagger\hat{a}_{\beta\mathbf{j}}^\dagger\hat{a}_{\alpha\mathbf{j}}\hat{a}_{\alpha\mathbf{j}}\right).
\end{equation}
Here, $\hat{n}_{\alpha\mathbf{j}}=\hat{a}_{\alpha\mathbf{j}}^\dagger\hat{a}_{\alpha\mathbf{j}}$ is the atomic number operator for flavor $\alpha$ at site $\mathbf{j}$. The indices $\alpha,\,\beta=x,\,y$, and the $\sum_{\langle{\bf ij}\rangle_\alpha}$ sums over nearest neighbors in the direction $\alpha$. The parameters are given by the overlap integrals; tunnelling amplitude
\begin{equation}\label{tunov}
t_{\alpha\beta}=-\int\,d\mathbf{r}\,w_{\alpha\mathbf{j}}(\mathbf{r})\hat{H}_\mathrm{sp}w_{\alpha\mathbf{j}+\mathbf{1}_\beta}(\mathbf{r}),
\end{equation}
where $\mathbf{j}+\mathbf{1}_\beta$ indicates the neighboring site of $\mathbf{j}$ in the direction $\beta$,
\begin{equation}
E_\alpha(t)=\int\,d\mathbf{r}\,w_{\alpha\mathbf{j}}(\mathbf{r})\left[-\nabla^2+V_\mathrm{lat}(\mathbf{r})\right]w_{\alpha\mathbf{j}}(\mathbf{r}),
\end{equation}
and the interaction strengths read
\begin{equation}\label{intov}
U_{\alpha\beta}=U_0\int\,d\mathbf{r}\,w_{\alpha\mathbf{j}}^2(\mathbf{r})w_{\beta\mathbf{j}}^2(\mathbf{r}).
\end{equation}
Note that we can choose the Wannier functions to be real. Without the LZ sweep, the onsite energies $E_\alpha(t)$ are assumed the same between the two orbitals, while by tuning $V_x$ and $V_y$ externally the onsite energy becomes time-dependent and we can drive the LZ sweep. In principle also the other parameters will be altered by tuning the amplitudes $V_x$ and $V_y$, but this effect will be much smaller than the change in $E_\alpha(t)$, and we can safely ignore such dependences. In the isotropic case and in the harmonic approximation, i.e. $w_{x\mathbf{j}}(\mathbf{r})\propto(x-\pi j_x)\exp\left(-(x-\pi j_x)^2/2\sigma-(y-\pi j_y)^2/2\sigma\right)$ with $\sigma$ the width and similarly for $w_{y\mathbf{j}}(\mathbf{r})$, the interaction strengths obey $U_{xx}=U_{yy}=3U_{xy}=3U_{yx}$. Therefor, we have defined $U=U_{xx}$ to parametrize all interaction terms. The effect of the trap, appearing as the last term in Eq. (\ref{kinterm}), where we have assumed that the trap varies minimally on the length scale of the lattice, i.e. the trap does not directly induce tunnelling between lattice sites. In this approximation we just replace $x_\mathbf{j}$ and $y_\mathbf{j}$ with the positions of the site $\mathbf{j}$. Finally we make a remark about the tunnelling coefficients $t_{\alpha\beta}$ which give the amplitude for an $\alpha$-orbital particle to tunnel in the $\beta$-direction. Due to the anisotropic shape of the orbitals, the tunnelling strength of say a $p_x$-orbital in the $x$-direction, $t_{xx}$, is not the same as for tunnelling in the $y$-direction, $t_{yx}$. Using the particular shapes of the Wannier functions on the $p$-bands it follows that $|t_{xx}|>|t_{yx}|$, and if we pick $t_{xx}<0$ we have that $t_{yx}>0$. For our purposes it is enough to consider the orbital symmetric situation $t_{xx}=t_{yy}\equiv t_1$ and $t_{xy}=t_{yx}\equiv t_2$ (corresponding to the isotropic lattice). Another consequence of the shapes of the $w_{\alpha{\bf j}}({\bf j})$'s is that tunnelling cannot accompany a change of orbital states (in general such tunnelings describe spin-orbit couplings or so called Dzyaloshinskii-Moriya processes), instead orbital states are only interchanged via scattering of two $p_x$-orbitals into two $p_y$-orbitals or vice versa, as seen from Eq.~(\ref{ocint}). 

Experimentally, atoms prepared in the ground state, the $s$-band, can be loaded into the $p$-bands via a two photon Raman pulse~\cite{bloch}, and there they would be let to relax into a $p$-band (quasi) ground state. Such relaxation is indeed very fast - a few tunnelling times~\cite{bloch,fernanda2}. This implies that it takes place on a much shorter time scale than other decay mechanisms. The lattice will initially be anisotropic ($V_x\neq V_y$) such that all atoms populate mainly one orbital type (the lattice anisotropy lifts the $p$-band degeneracy). The lattice is then gradually changed in order to realize the LZ sweep. The density of the cloud can be measured either via time-of-flight or by florescence measurements, which can work even at the single site level~\cite{ss}. 

\subsection{Many-body Landau-Zener Hamiltonian}\label{ssec21}
The idea of an interplay between physical mechanisms characterized by different time-scales means, in our case, that we consider a system of both internal ``spin'' (the two orbitals) and external spatial degrees of freedom. The idea is that if the external degrees of freedom are frozen we recover a regular (non-linear) LZ problem, while if the full system is considered the evolution of the external degrees of freedom taking place during the LZ sweep can affect the onsite LZ transition probabilities. As we will see below, the lengths of the onsite spinors are not fixed and my vary in time. In particular, the lengths give the number of atoms at that site. The change in the spinor length follows then from the fact that the atoms are mobile and can occupy different sites. We will demonstrate that this particle mobility renders very complex coupled evolution. We may note that this situation is qualitatively different from earlier studies of LZ lattice physics~\cite{nlzexp} where the occupation in the sites is fixed, and the presence of surrounding sites appears solely as a dephasing effect on the onsite problems.

In the following section were the results are presented we assume that a mean-field approximation is justified, i.e. the atom number on every single site is typically larger than ten. Even at the edge of the atomic cloud, where the particle number drops well below ten, we imagine that number fluctuations are still large and a mean-field description to be motivated. By virtue of the mean-field approach we make the coherent state ansatz, i.e.
\begin{equation}
|\Psi\rangle=|\psi_x,\psi_y\rangle=\bigotimes_\mathbf{j}|\psi_\mathbf{j}\rangle_\mathbf{j}=\bigotimes_\mathbf{j}|\psi_{x\mathbf{j}},\psi_{y\mathbf{j}}\rangle_\mathbf{j},
\end{equation}
where the $\mathbf{j}$-site state $|\psi_{x\mathbf{j}},\psi_{y\mathbf{j}}\rangle_\mathbf{j}$ is a two-mode coherent state
\begin{equation}\label{mfstate}
|\psi_{x\mathbf{j}},\psi_{y\mathbf{j}}\rangle_\mathbf{j}=\exp\!\left(\!-\frac{|\psi_{x\mathbf{j}}|^2\!+\!|\psi_{y\mathbf{j}}|^2}{2}\right)\!\!\!\sum_{n_x,n_y}\!\!\frac{\psi_{x\mathbf{j}}^{n_x}\psi_{y\mathbf{j}}^{n_y}}{\sqrt{n_x!n_y!}}|n_x,n_y\rangle_\mathbf{j}
\end{equation}
with $|n_x,n_y\rangle_\mathbf{j}$ a Fock state with $n_x$ and $n_y$ $p_x$- and $p_y$-orbital atoms at site $\mathbf{j}$ respectively. As coherent states we have $\hat a_{\alpha{\bf j}}|\Psi\rangle=\psi_{\alpha{\bf j}}|\Psi\rangle$ and $\langle\Psi|\hat n_{\alpha{\bf j}}|\Psi\rangle=|\psi_{\alpha{\bf j}}|^2$. Note that $|\psi_{x\mathbf{j}},\psi_{y\mathbf{j}}\rangle_\mathbf{j}$ represents the spinor at site $\mathbf{j}$ and from here it follows that its norm $n_\mathbf{j}=\sqrt{|\psi_{x\mathbf{j}}|^2+|\psi_{y\mathbf{j}}|^2}$ gives the number of particles at that particular site, i.e. $N_\mathrm{tot}=\sum_\mathbf{j}n_\mathbf{j}$ is the total number of particles. 

The mean-field Hamiltonian (or classical energy functional) is given by $H_\mathrm{mf}\left[\psi_{\alpha\mathbf{j}}\right]\equiv\langle\Psi|\hat{H}|\Psi\rangle$ where $\hat{H}$ is the second quantized Hamiltonian of Eq. (\ref{secondham}). After writing the Hamiltonian $\hat{H}$ on a normally ordered form, and using the fact $\hat{a}_{\alpha\mathbf{j}}|\Psi\rangle=\psi_{\alpha\mathbf{j}}|\Psi\rangle$ for any $\alpha=x,\,y$ and $\mathbf{j}$, it directly follows that the energy functional is simply obtained by replacing operators $\hat{a}_{\alpha\mathbf{j}}$ ($\hat{a}_{\alpha\mathbf{j}}^\dagger$) with $\psi_{\alpha\mathbf{j}}$ ($\psi_{\alpha\mathbf{j}}^*$, where $*$ represents complex conjugation). This gives us $H_\mathrm{mf}=H_0+H_\mathrm{dd}+H_\mathrm{oc}$ with
\begin{equation}
\begin{array}{lll}
H_0 & = & \displaystyle{-\sum_{\alpha,\beta}\sum_{\langle\mathbf{ij}\rangle_\alpha}t_{\alpha\beta}\psi_{\beta\mathbf{i}}^*\psi_{\beta\mathbf{j}}}\\ \\
& & \displaystyle{+\sum_\alpha\sum_\mathbf{j}\left[E_{\alpha}(t)+\frac{\omega^2}{2}\left(x_\mathbf{j}^2+y_\mathbf{j}^2\right)\right]|\psi_{\alpha\mathbf{j}}|^2},
\end{array}
\end{equation}
\begin{equation}
H_\mathrm{dd}=\frac{U}{2}\sum_\alpha\sum_\mathbf{j}|\psi_{\alpha\mathbf{j}}|^4+\frac{U}{3}\sum_{\alpha\beta,\alpha\neq\beta}\sum_\mathbf{j}|\psi_{\alpha\mathbf{j}}|^2|\psi_{\beta\mathbf{j}}|^2,
\end{equation}
\begin{equation}\label{oc}
H_\mathrm{oc}=\frac{U}{6}\sum_{\alpha\beta,\alpha\neq\beta}\sum_\mathbf{j}\left[\left(\psi_{\alpha\mathbf{j}}^*\psi_{\beta\mathbf{j}}\right)^2+\left(\psi_{\beta\mathbf{j}}^*\psi_{\alpha\mathbf{j}}\right)^2\right].
\end{equation}
The onsite energies $E_\alpha(t)$ are given by $E_x(t)=-\lambda t$ and $E_y(t)=\lambda t$ with $\lambda$ ($>0$) the velocity of the LZ sweep. Throughout, the interaction is taken to be repulsive, $U>0$. 

The mean-field equations of motion can be derived from the Hamilton's equations $\partial\psi_{\alpha\mathbf{j}}/\partial t=-\partial H/\partial\psi_{\alpha\mathbf{j}}^*$. The resulting equations (i.e. discrete Gross-Pitaevskii equations) become~\cite{fernanda1}
\begin{equation}\label{mfeom}
\begin{array}{l}
\begin{array}{lll}
\displaystyle{i\frac{\partial\psi_{x\mathbf{j}}}{\partial t}} & = & \displaystyle{-t_1\!\left(\psi_{x\mathbf{j+1}_x}\!+\!\psi_{x\mathbf{j-1}_x}\right)-t_2\!\left(\psi_{x\mathbf{j+1}_y}\!+\!\psi_{x\mathbf{j-1}_y}\right)}\\ \\
& & \displaystyle{+\frac{\omega^2}{2}\left(x_\mathbf{j}^2+y_\mathbf{j}^2\right)\psi_{x\mathbf{j}}+\lambda t\psi_{x\mathbf{j}}}\\ \\
& & +\displaystyle{U\left(|\psi_{x\mathbf{j}}|^2+\frac{2}{3}|\psi_{y\mathbf{j}}|^2\right)\psi_{x\mathbf{j}}+U\frac{2}{3}\psi_{y\mathbf{j}}^2\psi_{x\mathbf{j}}^*},
\end{array}
\\ \\
\begin{array}{lll}
\displaystyle{i\frac{\partial\psi_{y\mathbf{j}}}{\partial t}} & = & \displaystyle{-t_1\!\left(\psi_{y\mathbf{j+1}_y}\!+\!\psi_{x\mathbf{j-1}_y}\right)-t_2\!\left(\psi_{y\mathbf{j+1}_x}\!+\!\psi_{y\mathbf{j-1}_x}\right)}\\ \\
& & \displaystyle{+\frac{\omega^2}{2}\left(x_\mathbf{j}^2+y_\mathbf{j}^2\right)\psi_{y\mathbf{j}}-\lambda t\psi_{y\mathbf{j}}}\\ \\
& & +\displaystyle{U\left(|\psi_{y\mathbf{j}}|^2+\frac{2}{3}|\psi_{x\mathbf{j}}|^2\right)\psi_{y\mathbf{j}}+U\frac{2}{3}\psi_{x\mathbf{j}}^2\psi_{y\mathbf{j}}^*}.
\end{array}
\end{array}
\end{equation}
The above expressions make clear that $\psi_{\alpha\mathbf{j}}$ couples to its own complex conjugate $\psi_{\alpha\mathbf{j}}^*$. This peculiar coupling of the order parameter stems from the orbital changing term $\hat{H}_\mathrm{oc}$ given in Eqs.~(\ref{ocint}) and (\ref{oc}). Normally for Gross-Pitaevskii realizations appearing in atomic physics, the order parameter couples only to its density $|\psi|^2$ and not to $\psi^*$ alone~\cite{pethick,comspin}. 

The coherent state amplitudes $\psi_{\alpha\mathbf{j}}$ should be seen as the superfluid order parameter; the atoms are assumed condensed and $|\psi_{\alpha\mathbf{j}}|^2$ gives the number of atoms of flavor $\alpha$ in site $\mathbf{j}$. It should be remembered, however, that the full atomic density, taking the spatial dependences of the Wannier functions into account, is
\begin{equation}
P({\bf r})=\sum_{\alpha=x,y}\sum_{\bf j}|\psi_{\alpha{\bf j}}|^2w_{\alpha{\bf j}}^2({\bf r}).
\end{equation}
Nevertheless, in the following we will talk about the atomic density as the occupations of the different orbital states $|\psi_{\alpha{\bf j}}|^2$, and the total density at site ${\bf j}$
\begin{equation}
Q_{\bf j}=|\psi_{x{\bf j}}|^2+|\psi_{y{\bf j}}|^2,
\end{equation}
where as mentioned earlier, $|\psi_{\alpha{\bf j}}|^2$ gives the number of $p_\alpha$-orbital atoms in site ${\bf j}$.

\subsection{Multiple time-scale many-body Landau-Zener problem}\label{ssec22}
The presence of a trap is crucial for the system evolution during the LZ sweep. At first, this may seen as strange since the trap shifts the energies of the two orbitals equally within a single site. In another language, it seems to be a local change of an effective chemical potential. But as we will explain next, this is indeed {\bf not} a local density approximation, which derives from the anisotropic properties of the tunnelling, i.e. $|t_1|>|t_2|$.

The tunnelling part of the Hamiltonian drives the kinetics of particles within the lattice. Hence, the coefficients $t_1$ and $t_2$ can be seen as inverse effective masses for the particles. Since $|t_1|>|t_2|$, a $p_x$-orbital particle is `heavier' in the $y$-direction than in the $x$-direction, and oppositely for a $p_y$-orbital particle. Thus, an initially localized single $p_x$-orbital particle will diffuse more rapidly in the $x$-direction than in the $y$-direction. Furthermore, the effective mass is negative in the $x$-direction so the particle actually maintain both a `particle' and a `hole' character. Now, if such a single $p_x$-orbital particle is confined in a harmonic potential (atomic trap), its ground state wave function, $\psi_{x{\bf j}}$, is Gaussian in both directions, but its width in the $x$-direction is larger than in its $y$-direction. Naturally, the opposite holds true for a $p_y$-orbital particle. Interaction will couple the two orbital states, but as long as the interaction is not too strong (that is, we are not in the Thomas-Fermi regime~\cite{pethick}) the ground state of $N_\mathrm{tot}$ particles will not be polar symmetric even if the harmonic potential is isotropic~\cite{fernanda1}. That means that at sites at the boundary of the particle distribution, either $p_x$- or $p_y$-orbital particle densities will dominate. This demonstrates that the model goes beyond the local density approximation - the onsite potential strength does not uniquely determine the particle density (chemical potential) locally. 

How does this intrinsic anisotropy affect the LZ driving? Starting with say $E_x\ll E_y$ all particles will reside in the $p_x$-orbitals and the ground state particle distribution will be elongated in the $x$-direction. For $E_y\ll E_x$, on the other hand, the corresponding distribution will instead be elongated in the $y$-direction and all particles will populate $p_y$-orbitals. This observation will be explicitly demonstrated in the next Section. When we drive the LZ transition adiabatically it means that at every occupied site particles swap from $p_x$- to $p_y$-orbitals. Simultaneously, if we are to remain in the global ground state the external shape of the particle distribution must also change (otherwise we pay a price in potential energy); the squeeze-shaped atomic density should be rotated by 90 degrees. At every populated site a non-linear LZ transition is realized, but in addition, the sites are coupled and particles can hop between them. The onsite LZ transition occurs on some characteristic time $\tau_\mathrm{intra}$ that will depend on the sweep velocity $\lambda$ and the interaction strength $U$, but also on the tunnelings $t_1$ and $t_2$ (as explained in the next paragraph). The extrinsic dynamics during the LZ transition is characterized by some time $\tau_\mathrm{inter}$ which depends predominantly on the tunnelling amplitudes. Of course, this is a very simplified picture of the full coupled system, but it gives an idea of the complex dynamics. It follows that performing an adiabatic sweep would mean that $\lambda^{-1}\gg\tau_\mathrm{intra},\,\tau_\mathrm{inter}$. Physically, if we have a macroscopic number of particles in our lattice the above scenario implies that we need to achieve a macroscopic current of particles within the lattice. For physically relevant parameters one would therefore expect that $\tau_\mathrm{inter}>\tau_\mathrm{intra}$. Then, if $\tau_\mathrm{inter}>\lambda^{-1}>\tau_\mathrm{intra}$ the onsite LZ transitions can be adiabatic while the overall particle distribution becomes excited in terms of collective vibrations. 

A qualitative argument to explain the difference in time-scales, $\tau_\mathrm{intra}$ and $\tau_\mathrm{inter}$, goes as follows. We first note that the single site problem can be mapped onto a Lipkin-Meshkov-Glick LZ problem~\cite{jonas4}.  Due to the non-linearity appearing in the mean-field model, so called swallow-tail loops of the adiabatic energies form~\cite{nlz,pethick2}. Interestingly, in the present Lipkin-Meshkov-Glick system these loops are always present and they makes the gap between the lowest adiabatic energies to vanish. This would suggest that the intra time-scale $\tau_\mathrm{intra}$ should go to infinity. However, the present lattice model cannot be understood from single site evolution, and the fact that the sites are couple implies that an effective gap opens up between the lowest onsite adiabatic energies. This gap is of the order of the tunnelings $t_{1,2}$. In particular, the gap size does not depend, to lowest order, on the system size since the number of neighbors is fixed. On the other hand, the inter time-scale $\tau_\mathrm{inter}$ will depend on the total number of sites (the single particle tight-binding model is gapless), and in particular this gap vanishes in the thermodynamic limit. Thus, for sufficiently large lattices the slow time-scale will unambiguously be the inter site one.

\section{Numerical results}\label{mssec}
In this section the idea is to demonstrate numerically what we previously argued, namely that the coupling between internal and external evolution leads to very complex full system dynamics where in particular the Franck-Condon type physics alters the LZ transition. 

The many-site problem of Eq.~(\ref{mfeom}) is solved using the split-operator method~\cite{so}. This allows us to both consider time-dependent problems as well as extracting the ground state by simply propagating some initial state in imaginary time, $t\rightarrow-it$. For simplicity, the full state $|\Psi(t)\rangle$ will be normalized to unity (instead of the total number of atoms $N_\mathrm{tot}$). Of special interest for us is the atomic imbalance
\begin{equation}
Z_\mathrm{tot}(t)=\sum_\mathbf{j}Z_\mathbf{j}(t)=\sum_\mathbf{j}\left[|\psi_{x\mathbf{j}}(t)|^2-|\psi_{y\mathbf{j}}(t)|^2\right].
\end{equation} 
Note that if the onsite imbalance would be normalized, $Z_\mathbf{j}(t)\rightarrow Z_\mathbf{j}(t)/(|\psi_{x\mathbf{j}}(t)|^2+|\psi_{y\mathbf{j}}(t)|^2)$, it directly relates to the LZ transfer probability introduced in the introduction. It should be clear that an adiabatic LZ sweep in the full lattice system would imply that $Z_\mathbf{tot}(-\infty)=1\rightarrow Z_\mathrm{tot}(+\infty)=-1$. However, as will be discussed further, reaching $Z_\mathrm{tot}(+\infty)=-1$ is not a guarantee for total adiabatic evolution.

\begin{figure}[h]
\centerline{\includegraphics[width=8cm]{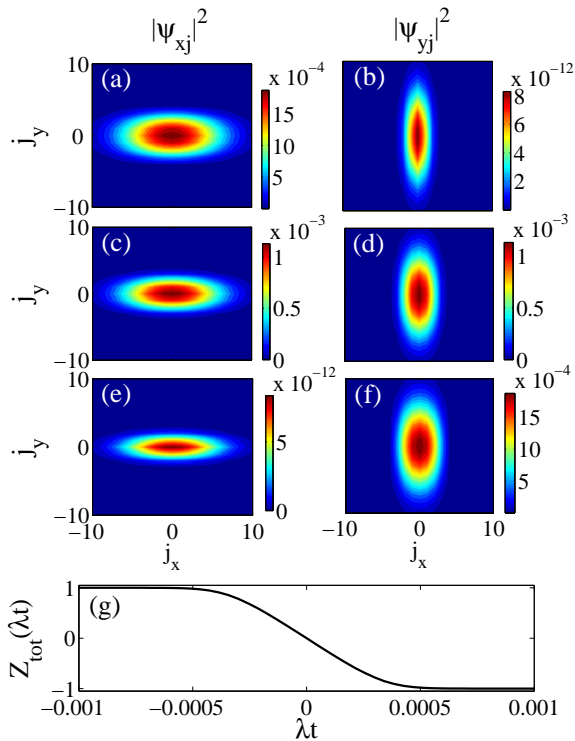}}
\caption{(Color online) Ground state distributions $|\psi_{\alpha{\bf j}}|^2$ (a)-(f) of the two $p_x$- (left) and $p_y$-orbitals (right) for different `detunings'; $\lambda t=-0.001$ (a) and (b), $\lambda t=0$ (c) and (d), and $\lambda t=0.001$ (e) and (f). The lower plot (g) shows the total atomic imbalance $Z_\mathrm{tot}(\lambda t)$. Apart from the regime $\lambda t\in\pm5\times10^{-4}$, the system ground state populates approximately only a single onsite orbital state (full polarization). The dimensionless parameters used are $\omega=0.003$, $t_1=-0.09$, $t_2=0.0045$, and $U=0.38$. The latter three numerical values correspond to an optical lattice with an amplitude of 17 recoil energies which is chosen to be an experimentally relevant situation~\cite{bloch}. The trap frequency is such that approximately a few hundred lattice sites are populated. } \label{fig7}
\end{figure}

\subsection{Ground state properties}
Before entering into the full time-dependent problem, let us look at the ground state properties of the time-independent system, i.e. $\lambda t$ is taken as a external parameter that fixes the detuning between the two orbitals. This will serve as a verification of the general argumentation put forward in Subsec.~\ref{ssec22}, and visualize the idea behind the Franck-Condon mechanism in our model. Given already in Eq.~(\ref{mfstate}), at the mean-field level, every site hosts a (non-normalized) qubit (spin-1/2 particle) characterized by a state  $|\psi_\mathbf{j}\rangle_\mathbf{j}=|\psi_{x\mathbf{j}},\psi_{y\mathbf{j}}\rangle=\left[\psi_{x\mathbf{j}}\,\,\psi_{y\mathbf{j}}\right]^T$ which alternatively can be represented by a Bloch vector $\mathbf{R}_\mathbf{j}=(J_{x\mathbf{j}},J_{y\mathbf{j}},J_{z\mathbf{j}})=(2\Re(\psi_{x\mathbf{j}}^*\psi_{y\mathbf{j}})\,,\,2\Im(\psi_{x\mathbf{j}}^*\psi_{y\mathbf{j}})\,,\,|\psi_{x\mathbf{j}}|^2-|\psi_{y\mathbf{j}}|^2)$. The length of the Bloch vector gives the (scaled) onsite particle occupation, and the $z$-component is nothing but the onsite particle imbalance. The LZ parameter $\lambda t$ acts as an external field which tries to align the onsite spins in the $z$-direction. Thus, for large $|\lambda t|$ we have $\mathbf{R}_\mathbf{j}/|\mathbf{R}_\mathbf{j}|=(0,0,\pm1)$, i.e. all the spins point either towards the north or the south pole on the Bloch sphere. This is conveniently called the polarized state. For zero field/detuning, $\lambda t=0$, and in the absence of a trap $\mathbf{R}_\mathbf{j}/|\mathbf{R}_\mathbf{j}|=(0,\pm1,0)$~\cite{girvin}. More precisely, due to the non-zero tunneling terms $t_1$ and $t_2$ (and their relative signs) and the character of the interaction terms, the full system organizes in an anti-ferromagnetic state with the spins alternating between pointing in the positive/negative $y$-direction between neighboring sites~\cite{girvin,fernanda1}. This observation suggests that in the thermodynamic limit there should occur an Ising type phase transition between these two phases.

Once the trap is included, as we have already argued, the densities of the two atomic orbitals becomes elongated despite the fact that the trap is isotropic. This is demonstrated in Fig.~\ref{fig7}. In the upper plots (a)-(f) we give examples of the ground state $p_x$-density (left plots) and $p_y$-density (right plots) for various $\lambda t$. More precisely, the plot gives the populations $|\psi_{\alpha{\bf j}}|^2$ of the two orbitals $w_{\alpha{\bf j}}({\bf r})$. Far from resonance, i.e. when $\lambda t=0$, we see that predominantly only $p_x$- or $p_y$-orbitals are populated (note the colourbars), and consequently the system is polarized in the $z$-direction. This is in agreement with interpreting $\lambda t$ as a field strength. For zero field, $\lambda t=0$ (c) and (d), the two distributions are identical but rotated 90 degrees. Thus, the full particle distribution including both orbitals is still not polar symmetric for $\lambda t=0$. In the lower plot (g) we show the total imbalance $Z_\mathrm{tot}$ for the whole lattice. Most interestingly, we find a non-vanishing regime around $\lambda t=0$ where a mixing of the two orbitals exists. We have numerically verified that in the thermodynamic limit, i.e. increasing the number of atoms while lowering the interaction strength $U$, the crossovers seen around $\lambda t\approx\pm0.0005$ become more sharp and finally they turn into a proper continuous phase transition. This is a transition between a polarized phase $Z_\mathrm{tot}=\pm1$ and a symmetry broken anti-ferromagnetic phase (mentioned above). This transition is of the Ising type~\cite{sachdev}, and it is interesting to notice that in the literature of cold atoms on the $p$-band of optical lattices~\cite{pband1} this transition in the superfluid regimes has been overlooked.

\subsection{Landau-Zener problem}
The discussed interplay between inter- and intra-site dynamics while driving the system through the LZ transition should be understood from Fig.~\ref{fig7}. In an adiabatic transition, the full particle distribution should go from the first to the third row of Fig.~\ref{fig7}, at the same time as the particles within each site are transferred from $p_x$-orbitals to $p_y$-orbitals (as in Fig.~\ref{fig7} (g)). Numerically we always consider a finite time sweep and thereby the ground state will always contain at least a small fraction of atoms in both orbital states. From $Z_\mathrm{tot}$ we have the amount of intrinsic excitations $P_\mathrm{iex}=(1-Z_\mathrm{tot})/2$. As already mentioned, this quantity is not capable of characterizing non-adiabaticity in the lattice since both inter- and intra-site excitations can exist. One direct measurement of how adiabatic the driving is would be to consider the instantaneous energy of the system and compare it to the corresponding ground sate energy. There is, however, a problem with such a measure in our model. Going back to the equations of motion (\ref{mfeom}) it is clear that what drives the transition is the last term which stems from the orbital changing interaction (\ref{oc}). If we start with all atoms residing in the $p_x$-orbital state it means that this coupling term vanishes identically as it is proportional to $\psi_{y\mathbf{j}}^2$. This is a result of considering a mean-field approximation; in a true system quantum fluctuations would `kick-off' the transition even with no $p_y$-orbital atoms initially. In other words, in this mean-field analysis we need to initially populate the $p_y$-orbitals in order to see any transition at all. Consequently, the system is thereby automatically in an excited state to begin with. An alternative approach would be to add a stochastic noise term to the equations of motion (\ref{mfeom}), but this implies that we need to perform sample averaging and a considerable slow down in the computations. Thus, we omit such an approach and it is also believed that the qualitative results would not change from the ones presented below. 

To understand the non-adiabatic excitations we will introduce the widths of the $p_y$-distribution
\begin{equation}
\Delta_y\alpha^2=\frac{\langle\psi_{y}|\hat{\alpha}^2|\psi_y\rangle}{\langle\psi_y|\psi_y\rangle}-\frac{(\langle\psi_y|\hat{\alpha}|\psi_y\rangle)^2}{(\langle\psi_y|\psi_y\rangle)^2},
\end{equation}
where $\hat{\alpha}$ ($\alpha=x,\,y$) is the discrete position operator. From the above widths we define the squeezing measure
\begin{equation}\label{squeeze}
F_y(t)=\frac{\Delta_yy^2}{\Delta_yx^2}
\end{equation}
which tells how elongated the $p_y$-distribution is; $F_y(t)=1\rightarrow$ no squeezing, $F_y(t)<1\rightarrow$ squeezing in the $y$-direction, and $F_y(t)>1\rightarrow$ squeezing in the $x$-direction. If the LZ sweep is adiabatic we have that at the final time $t_f$ $F_y(t_f)>1$ and moreover $F_y(t)$ should be time-independent for large times. Variations in $F_y(t)$ derive mainly from non-adiabatic excitations in terms of vibrations in the particle distribution (phonons), and we thereby introduce $\delta F_y(t)$ as the time-variance of the squeezing parameter at time $t$. In the following, $P_\mathrm{iex}$ and $\delta F_y(t)$ will be our rough measures of intra- and inter-well non-adiabatic corrections respectively. But it should be kept in mind that there is not a one-to-one relation between these quantities and the LZ induced excitations.

\begin{figure}[h]
\centerline{\includegraphics[width=8cm]{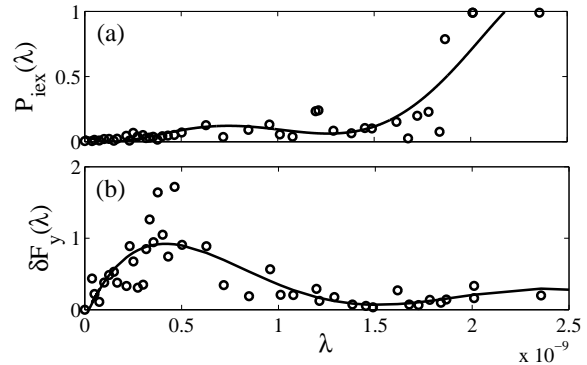}}
\caption{Intrinsic excitations $P_\mathrm{iex}(\lambda)$ (a) and variance of the squeezing parameter $\delta F_y(\lambda)$ (b). The open circles mark calculated values, while the solid line is a $5$'th order polynomial least square fit. Combining the results of the two plots, near full adiabatic population transfer is only encountered for $\lambda<10^{-10}$. The parameters are the same as for Fig.~\ref{fig7}, i.e. they correspond to a lattice amplitude of 17 recoil energies.} \label{fig8}
\end{figure}

In Fig.~\ref{fig8} we present numerical results from integrating Eq.~(\ref{mfeom}) for various sweep velocities $\lambda$. The integration interval $[t_i,t_f]$ is taken long enough such that the effective diabatic states $|\psi_x\rangle$ and $|\psi_y\rangle$ are approximately decoupled at $t_i$ and $t_f$ . The initial state is obtained from first finding the ground state (which will almost entirely populate the $p_x$-orbitals) for the given $\lambda t_i$. We then populate the $p_y$-orbitals with one percent by altering the initial state artificially where we take the $p_y$-orbital distribution $|\psi_{y{\bf j}}|^2$ to be the same as the one for the $p_x$-orbitals, $|\psi_{x{\bf j}}|^2$. By increasing the initial population in the $p_y$-orbitals the intrinsic evolution would in general be more adiabatic since the gap scales with $|\psi_y|^2$. However, this would also imply that the system would initially be more excited. In the figure we vary $\lambda$ while the remaining parameters are calculated from Wannier function overlap integrals, Eqs.~(\ref{tunov}) and (\ref{intov}), corresponding to an optical lattice with an amplitude of 17 recoil energies. This particular choice is meant to represent an experimentally relevant situation. The upper plot of Fig.~\ref{fig8} shows $P_\mathrm{iex}(t_f)$ as a function of $\lambda$, and the lower plot gives $\delta F_y(t_f)$ at the same instant. Due to long computational times, only a few values of $\lambda$ have been considered. The solid line is a fit of a fifth order polynomial to the calculated data.

An interesting question is whether $P_\mathrm{iex}(\lambda)$ can be assigned an exponential or a power-law dependence of $\lambda$. Characteristic for the standard LZ problem is its exponential dependence on the coupling strength $U$ and the sweep velocity $\lambda$ -- the transition probability is a smooth and monotonous function of both $U$ and $\lambda$. Furthermore, from the form of $P$ (given in the introduction) it follows that the result for small $\lambda$ (i.e. for adiabatic evolution) is non-perturbative. Extensions to multi-level problems~\cite{threelz1,bowtie,julienne}, many-body situations~\cite{jonas4,altland,fleischhauer,lmgcrit,orso}, and non-linear LZ transitions~\cite{nlz,lmglz,lmg2} have been considered. When the LZ model becomes non-linear both the exponential dependence and the smoothness of $P$ may be lost~\cite{nlz}. Such non-linear LZ problems typically arise in mean-field theories for single site LZ problems. It is particularly found that for strong enough non-linearity, adiabaticity cannot be achieved regardless of how slow the LZ sweep is~\cite{pethick2}. In addition, instead of an exponential dependence in the adiabatic regime the transition probability was found to obey a power-law dependence, i.e. $P_y\sim\lambda^\nu$ for some power $\nu$~\cite{nlz}. Power-law dependences have also been predicted in many-body LZ problems beyond the mean-field regime~\cite{jonas4,altland,fleischhauer}. The large fluctuations and the few data points of Fig.~\ref{fig8} make it impossible to extract any reliable power-law dependence of $P_\mathrm{iex}(t_f)$ in the adiabatic regime. It is found, however, that $P_\mathrm{iex}(t_f)$ shows a weak $\lambda$-dependence for $\lambda<2\times10^{-9}$, and then rapidly approach unity. This weak $\lambda$-dependence for small sweep velocities followed by a ``rapid'' change in the transition probability directly imply that trying to fit a curve $P(\lambda)=\exp(-C/\lambda)$ suggested by the LZ formula, for some fitting parameter $C$, would give a large discrepancy. For example, $P(\lambda)$ displays a rapid increase for small $\lambda$'s and then a slow saturation to its asymptotic value. Thus, it is clear that the transitions of the present model greatly differ from those of a single linear LZ problem.

\begin{figure}[h]
\centerline{\includegraphics[width=7cm]{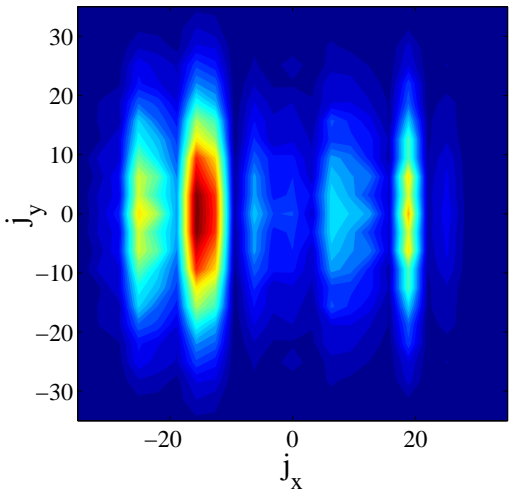}}
\caption{(Color online) The final (full) distribution $Q_{\bf j}(t_f)=|\psi_{x{\bf j}}(t_f)|^2+|\psi_{y{\bf j}}(t_f)|^2$ of the two orbitals in a situations where non-adiabatic excitations occurs mainly in the external degrees of freedom ($P_\mathrm{iex}=0.026$), i.e. in phononic vibrations of the particle distribution. The distribution is still squeezed in the $y$-direction despite the fact that almost only the $p_y$-orbital states are populated which clearly indicates that the transition takes place in the Franck-Condon regime. The sweep velocity $\lambda=1.68\times10^{-9}$ and the rest of the parameters are as in Fig.~\ref{fig7}. } \label{fig9}
\end{figure}

\begin{figure}[h]
\centerline{\includegraphics[width=8cm]{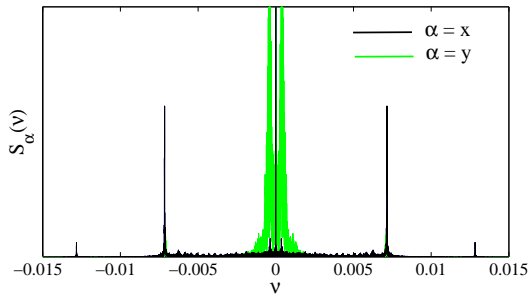}}
\caption{(Color online) The `spectral function' $S_\alpha(\nu)$ for the $x$ and $y$ vibrations of the full atomic distribution $Q_{\bf j}$. The fourier transform is taken in the interval preceding the LZ transition. The parameters are as in Fig.~\ref{fig9}. } \label{fig3}
\end{figure}

As demonstrated in Fig.~\ref{fig8} (b), the fluctuations in the squeezing measure reach a peak in the regime when intra-site excitations are still relatively low. Thus, as discussed earlier, the internal time-scale $\tau_\mathrm{intra}$ is shorter than the external one $\tau_\mathrm{inter}$, and non-adiabatic excitations predominantly occur as vibrational phonon modes of the particle distribution. Upon increasing $\lambda$, the fluctuations of $F_y(t_f)$ decrease and here it is actually found that the distribution has not performed any `rotation', i.e. the LZ sweep has mainly taken place onsite, or in other words this is deep in the Franck-Condon regime. In Fig.~\ref{fig9} we give an example (snapshot) of the final full distribution $Q_{\bf j}(t_f)=|\psi_{x{\bf j}}(t_f)|^2+|\psi_{y{\bf j}}(t_f)|^2$. Here $P_\mathrm{iex}=0.026$ implies good intra-site LZ transfer, but from the plot it is clear that external excitations in terms of particle vibrations are large. Such large quantum fluctuations are indeed common for many-body LZ problems~\cite{jonas4,altland}. The difference in this study compared to earlier ones is that we can characterize the excitations into two categories of internal and external, and thereby also quantify the types of excitations. The external excitations can be analyzed in terms of the fourier spectrum of some physical quantity $A(t)=\langle\hat{A}\rangle$. Here we define the `spectrum' as the fourier transform of the $x$- and $y$-`widths' of the full distribution $Q_{\bf j}$,
\begin{equation}\label{spectrum}
S_\alpha(\nu)\propto\int_{t_f}^\infty\,dt\,e^{i\nu t}\langle\hat{\alpha}^2\rangle,\hspace{1cm}\alpha=x,\,y.
\end{equation}  
Here, $t_f$ is the time when the LZ sweep stops and after that the system is evolving with a constant detuning $\lambda t_f$. Thus, $S_\alpha(\nu)$ is not exploring the evolution through the LZ transition, but only the vibrations in the full distribution caused be the LZ sweep. The results are displayed in Fig.~\ref{fig3} for both the directions. There are some common vibrational mode frequencies for both directions, $\nu=0,\,\pm0.0072,\,\pm0.0128$. The $y$-component also have two additional clear modes at $\nu=\pm0.0004$. These excitation frequencies are of the same order as the tunnelings $|t_1|$ ($=0.09$) and $t_2$ ($0.0045$) as well as the trap frequency $\omega$ ($=0.003$). Since $|t_1|$ determines the band width excitations occur within the band. In particular, the amplitude of the vibrational modes is much smaller than the typical energy gap to other bands. For a potential amplitude $V=17$ (in scaled units), we have the gap $\sqrt{2V}\sim6\gg\nu$. Thus, the characteristic energies of these phonon modes are at least two orders of magnitude smaller than excitations between bands within the lattice and as a result we conclude that the single-band approximation should be valid.

\begin{figure}[h]
\centerline{\includegraphics[width=7cm]{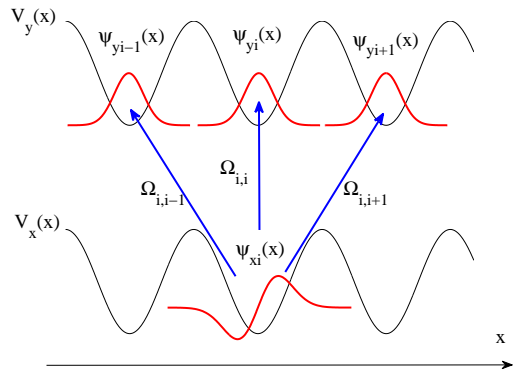}}
\caption{(Color online) The Franck-Condon mechanism. We imagine that initially the particle's state is $\psi_{xi}(x)$ (schematically represented by a $p_x$-orbital in the plot) and a coupling $\Omega$ between the $p_x$- and $p_y$-orbitals is turned on. If the coupling realizes a $\pi$-pulse transition, the particle will reside solely on the upper $p_y$-orbital branch after the turn-off of the coupling. In the Franck-Condon regime the state will be approximately $\psi_{yi}(x)$ (a $p_y$-orbital state at site $i$), i.e. the effective coupling strengths $|\Omega_{i,i}|\gg|\Omega_{i,i\pm1}|$ and the pulse is short such that tunnelings are negligible during the transfer. It is clear that for a pulse with a long duration the external evolution will affect the final state on the $p_y$-orbital branch.} \label{fig4}
\end{figure}

Let us end by a remark on how this model shows similarities to situations in molecular physics and in this sense can serve as a controllable model for studies of phenomena known from this field. The present LZ process is a kind of realization of Franck-Condon physics~\cite{fc}. The idea of the Franck-Condon mechanism is schematically described in Fig.~\ref{fig4}. The Franck-Condon principle plays an important role in molecular physics where the transition takes place between electronic states. It is used to understand internal molecular dynamics in pump-probe experiments. Here, the transition is between the orbital states, which belong to different Bloch bands, and the vibrational motion of the molecule is here replaced by particle motion within the lattice. One can imagine that by increasing the tunneling rates $t_1$ and $t_2$ it could be possible to be in a regime were $\tau_\mathrm{intra}\sim\tau_\mathrm{inter}$ and the Franck-Condon principle would not hold any longer. Such a situation might be difficult to achieve experimentally since in this regime the single-band and tight-binding approximations fail. However, there are probably ways to circumvent such issues, for example by considering non-separable lattices where the single-band approximation is much more easily fulfilled~\cite{hemmerich}. The tight-binding approximation is, in principle, not expected to be too crucial for the present analysis and consequently it is possible that the present system can work as a testbed for studies of Franck-Condon physics and its breakdown in a controlled manner. Moreover, here we analyze LZ transitions, but one could consider other schemes more similar to pump-probe methods, like Raman transitions~\cite{bloch} or lattices shaking~\cite{shake}.

\section{Conclusion}\label{seccon}
In this work we considered a LZ lattice problem at the mean-field level. While the present model could be experimentally realized in various types of systems, here we focused on one of cold bosonic atoms loaded into the $p$-bands of optical lattices. The novel feature of this system, appearing naturally on the $p$-band, derived from an interplay between intra-site LZ transitions and inter-site particle dynamics. The ability of particles to tunnel between lattice sites resulted in a much stricter constrain for adiabaticity. More precisely, an adiabatic evolution implied a macroscopic particle flow within the lattice. The non-adiabatic excitations appear as phonons making the particle distribution non-stationary. For physically relevant parameters it was found that the intra-site time-scale was much shorter than the inter-site time-scale which resulted in a Franck-Condon scenario.  In this cold-atom system, time-of-flight or single site addressing measurements would provide direct insight into both internal and external excitations created during the LZ sweep.

All results of this work assume a relatively large particle number per site (typically $>10$) where mean-field approximations start to give an accurate description of the physics. An interesting continuation would be to consider the opposite regime of a low filling and were strong correlations become dominant, e.g. in the insulating phase. As recently pointed out~\cite{fernanda2}, the physics of this system in the Mott insulator with unit filling is extremely rich and in particular realizes a $XY\!Z$ Heisenberg model, where the LZ sweep would represent a gradual change in the external field.  

\begin{acknowledgments}
The author acknowledges Fernanda Pinheiro for stimulating discussions and help with the numerical codes, and VR (Vetenskapsr\aa det) and KAW /the Knut and Alice Wallenberg foundation) for financial help.
\end{acknowledgments}


\begin{thebibliography}{999}
\bibitem{comspin} To achive this type of terms in spinor condensates one would need a significant large amplitude for scattering processes which convert atomic Zeeman levels. These are, however, often only on the order a few percent compared to other amplitudes.
\end{thebibliography}
\end{document}